\documentclass{moriond}

\def\Journal#1#2#3#4{{#1} {\bf #2}, #3 (#4)}

\def\EPJ{{\em Eur. Phys. J.} C}
\def\JHEP{\em J. High Energy Phys.}

\def\NIM{\em Nucl. Instrum. Methods}

\def\NPB{{\em Nucl. Phys.} B}
\def\PLB{{\em Phys. Lett.}  B}
\def\PR{\em Phys. Rep.}
\def\PRL{\em Phys. Rev. Lett.}
\def\PRD{{\em Phys. Rev.} D}

\def\ra{\rightarrow}

\def\be{\begin{equation}}
\def\ee{\end{equation}}
\def\bea{\begin{eqnarray}}
\def\eea{\end{eqnarray}}

\def\afb{A_{FB}}
\def\sinw{sin^{2}\theta_{W}}
\def\sineff{sin^{2}\theta^{l}_{eff}}
\def\cost{\cos\theta^*}
\newcommand{\Photo}{}

\begin{document}
\hspace{4.2in} \mbox{FERMILAB-CONF-14-311-E}
\vspace*{4cm}
\title{Z BOSON ASYMMETRY MEASUREMENTS AT THE TEVATRON}

\author{ B. QUINN \\
for the CDF and D0 Collaborations}

\address{Department of Physics and Astronomy, University of Mississippi, 
University, MS 38677, USA}

\maketitle\abstracts{
We present measurements of the forward-backward asymmetry ($\afb$) in dilepton 
pair decays of $Z$ bosons produced in $p\bar{p}$ collisions using the full 
Tevatron dataset.  The CDF experiment extracts a value for the effective weak 
mixing angle parameter $\sineff$ of $0.2315\pm0.0010$ from the $\afb$ 
distribution of dimuon events in 9.2 fb$^{-1}$ of integrated luminosity.  From 
dielectron events in 9.7 fb$^{-1}$ of data, the D0 experiment finds 
$\sineff = 0.23106\pm0.00053$, the world's most precise measurement of $\sineff$ 
from hadron colliders and with light quark couplings.}

\section{Introduction}

Drell-Yan lepton pairs~\cite{dy} are produced at the Tevatron through the 
reaction 
\begin{equation}
p\bar{p} \rightarrow Z/\gamma^* \rightarrow \ell^+\ell^-
\label{eq:DrellYan}
\end{equation}
at $\sqrt{s}=1.96$ TeV.  The angular distribution of these pairs is sensitive to 
the weak mixing angle through the vector coupling to the $Z$ boson, 
$g^f_V = I_3 - 2Q_f \sinw$.  This coupling is altered by weak radiative 
corrections of a few percent to give an effective weak mixing parameter, 
$\sineff$.

The angular distribution is measured in the Collins-Soper rest frame of the $Z$,
in which $\theta^*$ is defined as the angle of the $\ell^-$ relative to the 
direction of the incoming quark.~\cite{cs}  Events are categorized as forward (backward) if
$\cost \geq 0$ ($\cost < 0$).  After integrating over the azimuthal 
angle, the next to leading order QCD expression for the angular distribution 
becomes~\cite{mo}
\begin{equation}
dN/d\Omega \propto 1 + cos^2\theta^* + A_4\cost.
\label{eq:angdist}
\end{equation}
The forward-backward asymmetry is
\begin{equation}
\afb = \frac{\sigma^+ - \sigma^-}{\sigma^+ + \sigma^-} = \frac{3}{8}A_4,
\label{eq:Afb}
\end{equation}
where $\sigma^+$ and $\sigma^-$ are the forward and backward cross sections, 
respectively, and the $A_4$ parity violating term is sensitive to the weak 
mixing angle through $Z$ self-interference.

Both the CDF~\cite{CDF} and D0~\cite{D0} experiments employ general, multi-purpose detectors 
featuring excellent central tracking, calorimeter, and muon identification 
systems particularly relevant to these analyses and described in detail 
elsewhere.  CDF has recently published a measurement of $\sineff$ from $\afb$ in
dimuon events.~\cite{CDF_dimuon}  D0 has a new preliminary $\sineff$ result in the dielectron
channel.~\cite{D0_dielectron}  Both analyses are carried out in four steps:  measure $\afb$ from
dilepton events in bins of dilepton invariant mass ($M$), produce Monte Carlo 
(MC) templates of $\afb(M,\sinw)$, perform full corrections to data and 
simulation, and extract $\sineff$ by doing a $\chi^2$ comparison between the 
data and MC.

\section{CDF: $\sineff$ from dimuon events}

\begin{figure}
\includegraphics[width=0.5\linewidth]{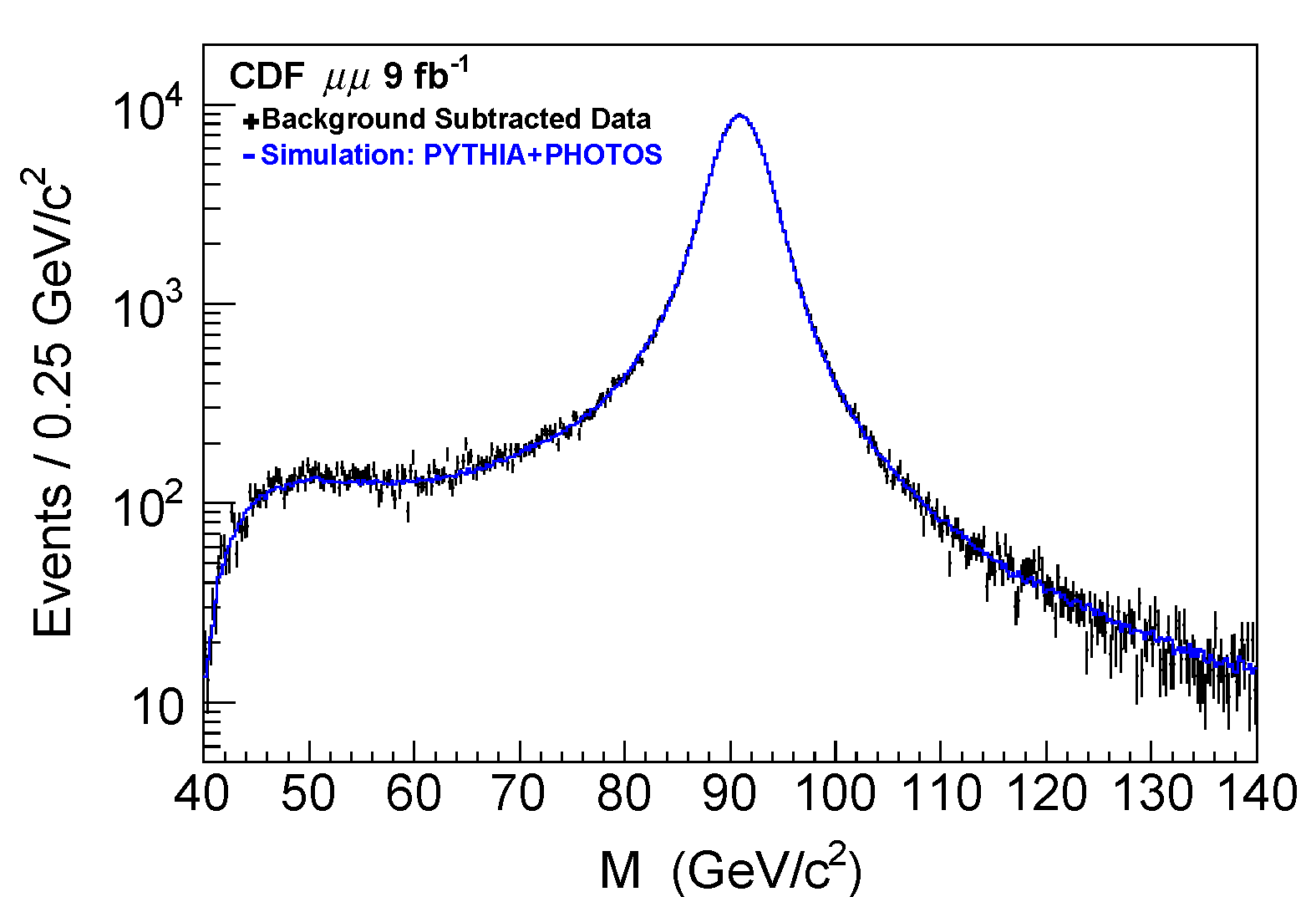}
\hfill
\includegraphics[width=0.5\linewidth]{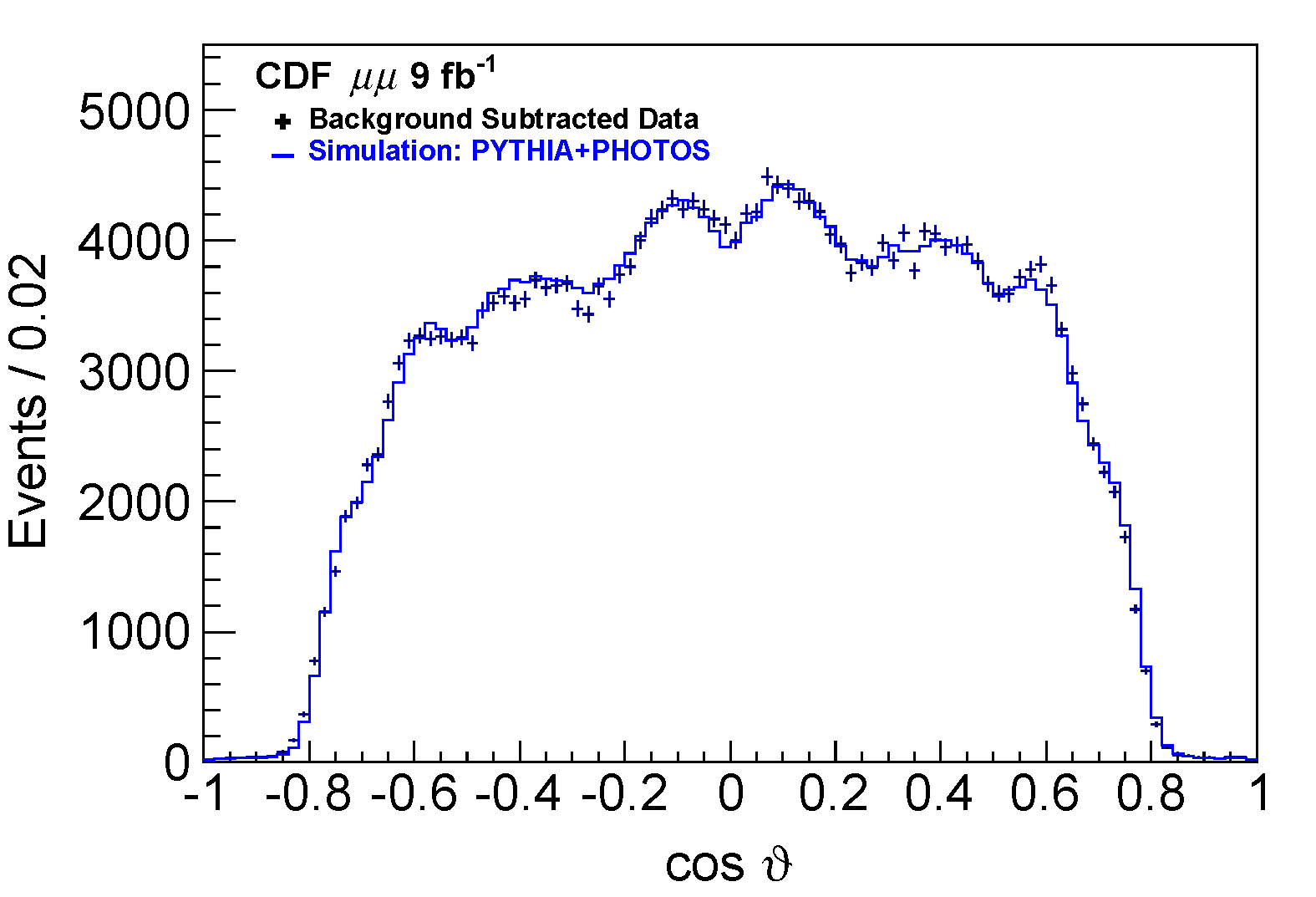}
\caption[]{CDF dimuon invariant mass (left) and Collins-Soper frame $\cost$ 
(right) distributions.}
\label{fig:CDFdist}
\end{figure}

CDF collected a sample of dimuon Drell-Yan pairs from the full dataset 
comprising 9.2 fb$^{-1}$ of integrated luminosity.  Selection cuts on the muons 
included the transverse momentum and rapidity requirements $p_T > 20$ GeV and 
$|y| < 1$, respectively, as well as $M > 50$ GeV.  A key part of the analysis is
a very precise momentum calibration using a method that tunes the data and
simulation to post-final state radiation (FSR) generator level distributions in 
64 individually calibrated bins of pseudorapidity and azimuthal angle.  Dimuon 
events with eleven different muon subdetector topologies of muon location were 
included in the sample.  Backgrounds from electroweak sources ($WW, WZ, ZZ, t\bar{t}, W+$jets, $Z\ra\tau^+\tau^-$) 
are estimated from MC to be $0.53\%$.  QCD backgrounds are determined from data 
to be a negligible $0.10\%$.  MC simulations were \textsc{pythia}-based,~\cite{pythia} using \textsc{cteq5l}~\cite{cteq5l} parton distribution functions (PDFs) and a \textsc{geant}~\cite{geant} simulation of the CDF detector.  After background subtraction, 276,623 events remained in the dimuon 
sample.  The data and MC distributions of $M$ and $\cost$ in the final sample are
shown in Fig.~\ref{fig:CDFdist}.

\begin{figure}
\includegraphics[width=0.5\linewidth]{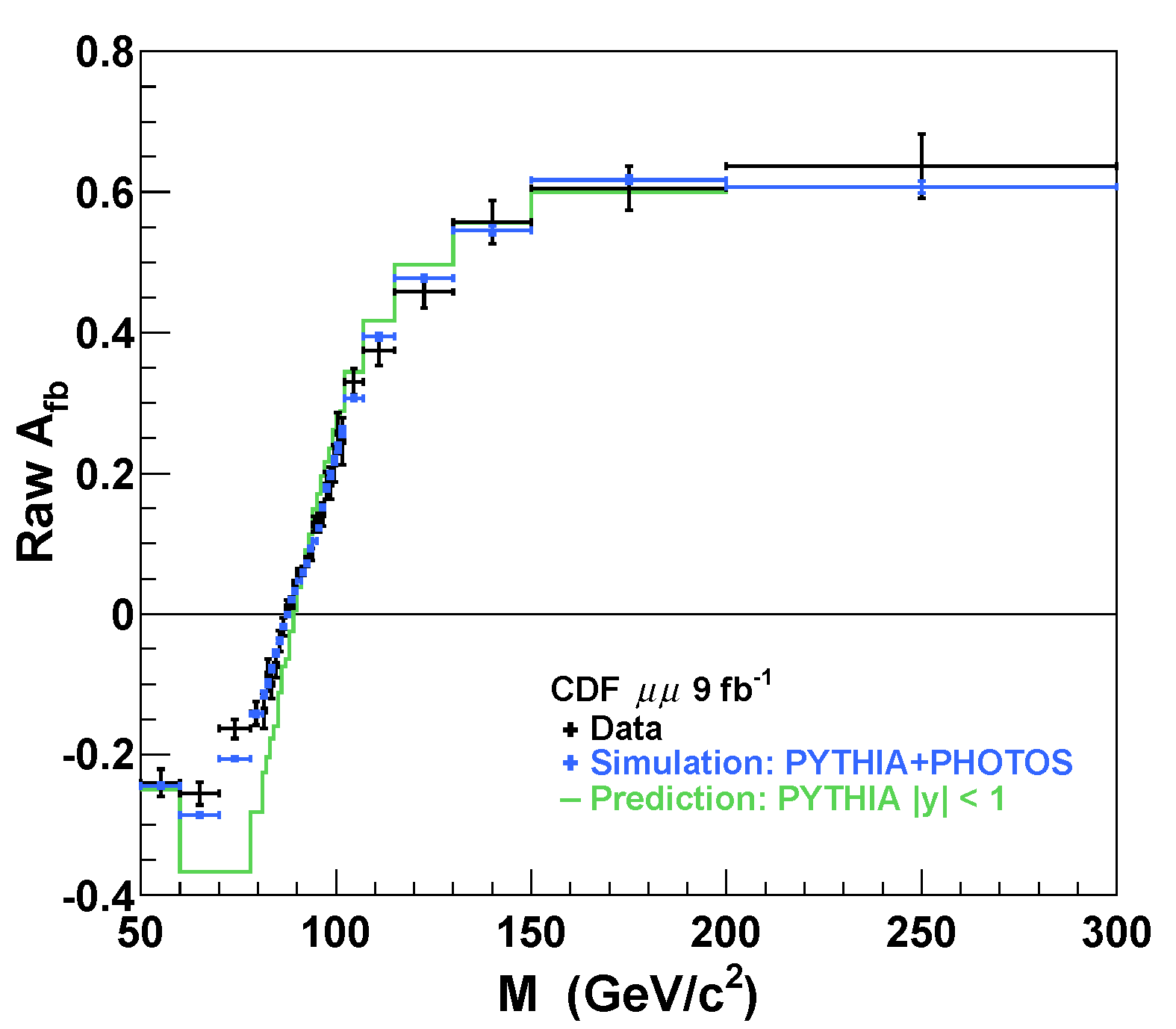}
\hfill
\includegraphics[width=0.5\linewidth]{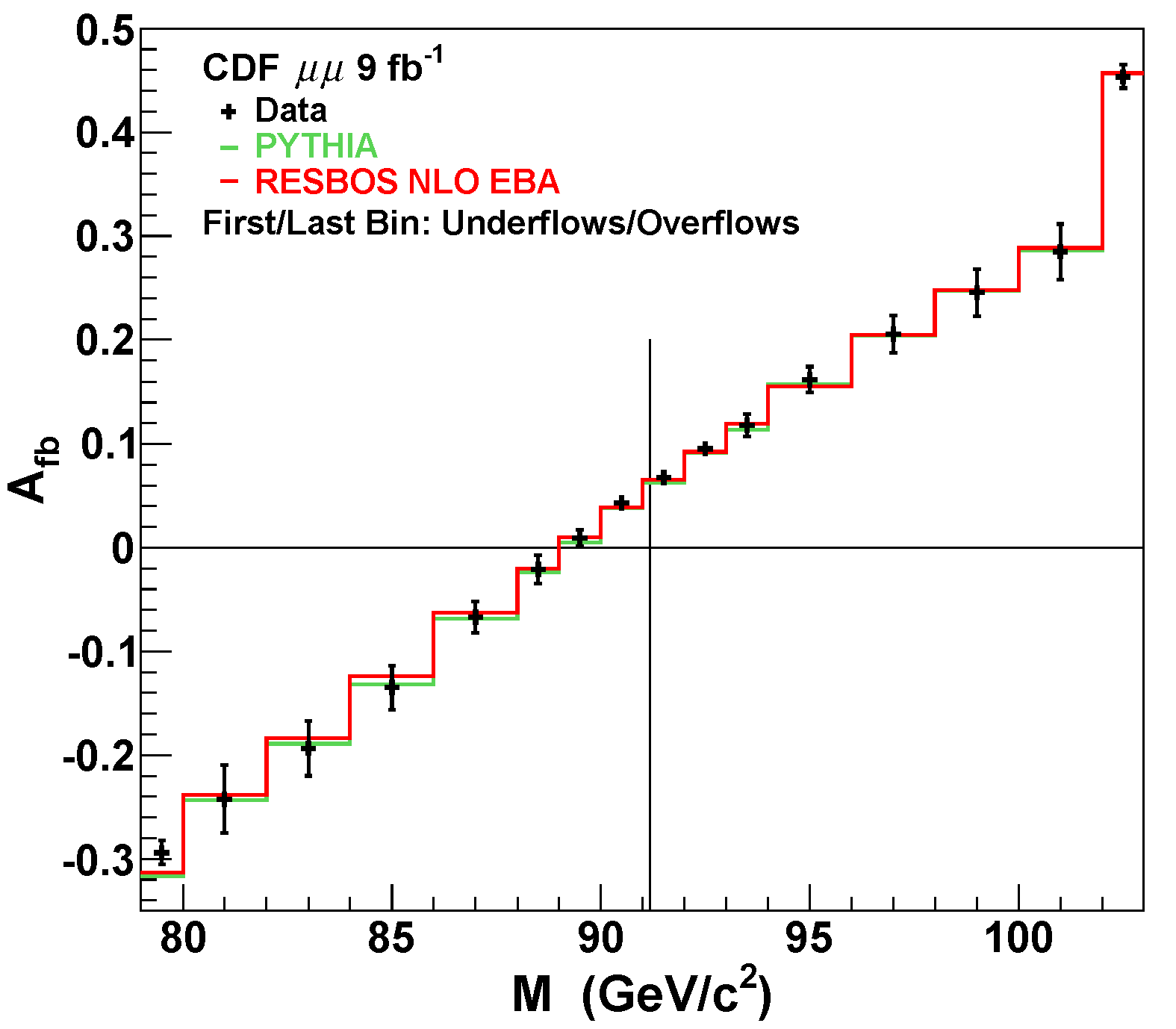}
\caption[]{CDF raw $\afb$ (left) and fully corrected $\afb$ (right) 
distributions.}
\label{fig:CDF_Afb}
\end{figure}

For this analysis, a traditional asymmetry measurement applying 
efficiency$\times$acceptance corrections to calculate cross sections (e.g. 
$\sigma^+ = N^+/(\epsilon A)^+)$ would be quite challenging.  It would require 
22 correction numbers for the 11 different muon topolgies, resulting in 
statistical limitations from this subdivision of the data sample.  An alternative
event weighting method~\cite{bodek} is employed which is equivalent to measuring $\afb$ in bins of 
$cos\theta^*$.  The asymmetry is calculated according to 
\begin{equation}
\afb(|\cost|) = \frac{\afb|\cost|}{1+\cos^2\theta^*+...},
\label{eq:angweight}
\end{equation}
assuming local $\epsilon A$ equivalence for forward and backward events and NLO 
QCD angular distribution.  This angular weighting results in more accurate 
measurements in bins of large $\cost$.  The binned measurements are then recast
into an unbinned weighted event sum
\begin{equation}
\afb = \frac{N^+_n - N^-_n}{N^+_d + N^-_d}, 
\label{eq:unbinned}
\end{equation}
where the $n,d$ subscripts refer to numerator and denominator event weights which
remove the angular dependencies from Eq.~\ref{eq:angweight} while preserving 
measurement accuracy at each $\cost$.  This is equivalent to performing a 
maximum likelihood fit, and delivers an expected $20\%$ reduction in uncertainty.
The raw $\afb$ distribution seen in Fig.~\ref{fig:CDF_Afb}(left) must then be 
unfolded to correct for resolution smearing and QED FSR effects.  Final 
bin-by-bin $2^{nd}$ order bias corrections are then applied to account for 
limited rapidity coverage and detector non-uniformities, yielding the fully 
corrected $\afb$ distribution shown in Fig.~\ref{fig:CDF_Afb}(right).

The $\afb$ measurement is then compared to $\afb$ templates calculated at 
different values of $\sinw$.  Three sets of templates are used, each with a 
different Enhanced Born Approximation (EBA) calculation (NLO \textsc{resbos}~\cite{resbos} , NLO \textsc{powheg-box}~\cite{powheg} 
, and LO tree), and all producing consistent results.  The \textsc{resbos} comparison 
gives the smallest $\chi^2$ and is selected as the default template set.  The 
dominant systematic uncertainty comes from the PDF uncertainty.  Smaller uncertainties include those on EBA calculation, 
backgrounds, and momentum and QCD scales.  All systematic uncertainties are much smaller than the statistical uncertainty of the measurement.  The $\chi^2$ fit to the \textsc{resbos} 
templates gives the value $\sineff = 0.2315\pm0.0009_{stat}\pm0.0004_{syst}$.  
Using an on-shell renormalization scheme in a Standard Model (SM) context, this
result is converted to $\sinw = 0.2233\pm0.0008_{stat}\pm0.0004_{syst}$, from 
which is obtained an indirect measurement of the $W$ boson mass, 
$M_W = 80.365\pm0.043_{stat}\pm0.019_{syst}$.

\section{D0: $\sineff$ from dielectron events}

Dielectron events were selected from D0's full dataset of 9.7 fb$^{-1}$.  
Electrons found in the central (CC) and endcap (EC) caolrimeters with $p_T > 25$ 
GeV and $M > 50$ GeV were accepted.  Compared to the previous D0 5 fb$^{-1}$ 
publication,~\cite{D0_5fb} an increase in statistics $85\%$ beyond luminosity scaling was 
achieved by extending the pseudorapidity range to $|\eta| < 1$ and 
$1.5 < |\eta| < 3.2$, including events with both electrons in the endcaps and 
those with electrons near calorimeter module boundaries, and improving track 
reconstruction.  The final sample consisted of 560,267 events with low QCD and 
negligible electroweak backgrounds totaling $0.4\%$ for central and $<4\%$ for 
endcap events.  The MC samples were generated with \textsc{pythia}~\cite{pythia} using \textsc{cteq6l1}~\cite{cteq6l1} PDFs and 
a \textsc{geant}-based~\cite{geant} simulation of the D0 detector. 

\begin{figure}
\includegraphics[width=0.5\linewidth]{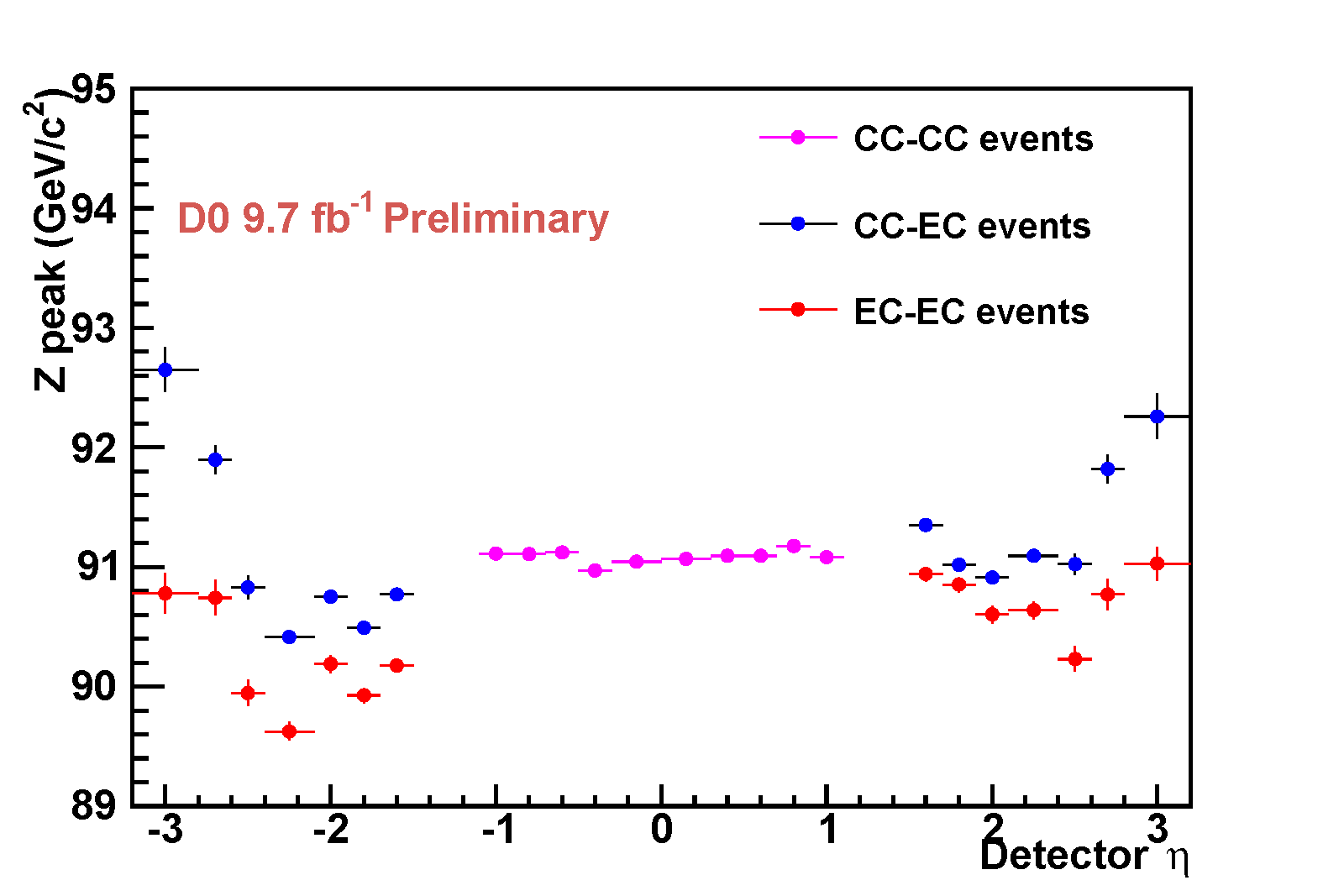}
\hfill
\includegraphics[width=0.5\linewidth]{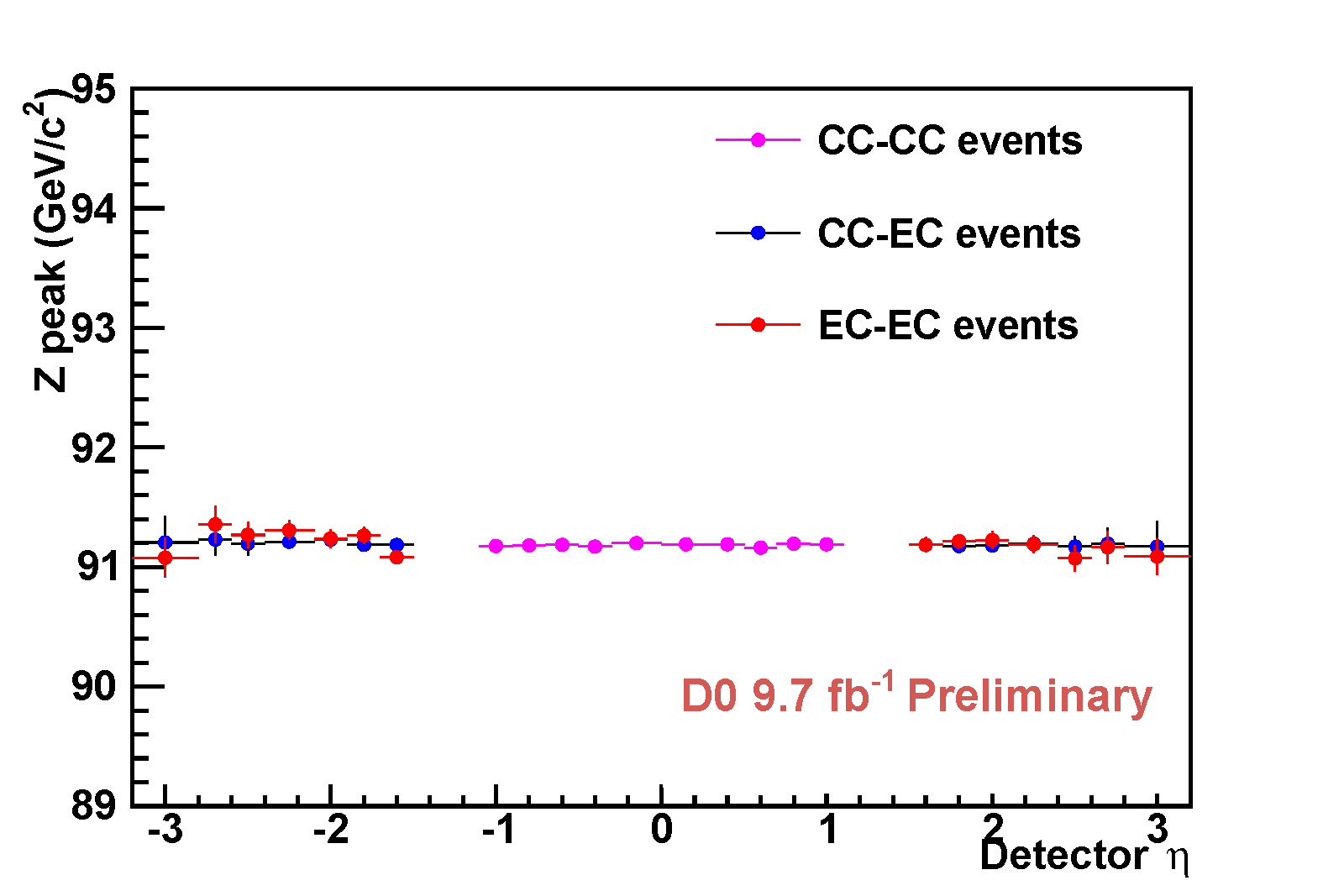}
\caption[]{D0 comparison of $Z$ mass peak position as a function of detector
pseudorapidity before (left) and after (right) binned energy scale calibration.}
\label{fig:escale}
\end{figure}

\begin{figure}
\includegraphics[width=0.5\linewidth]{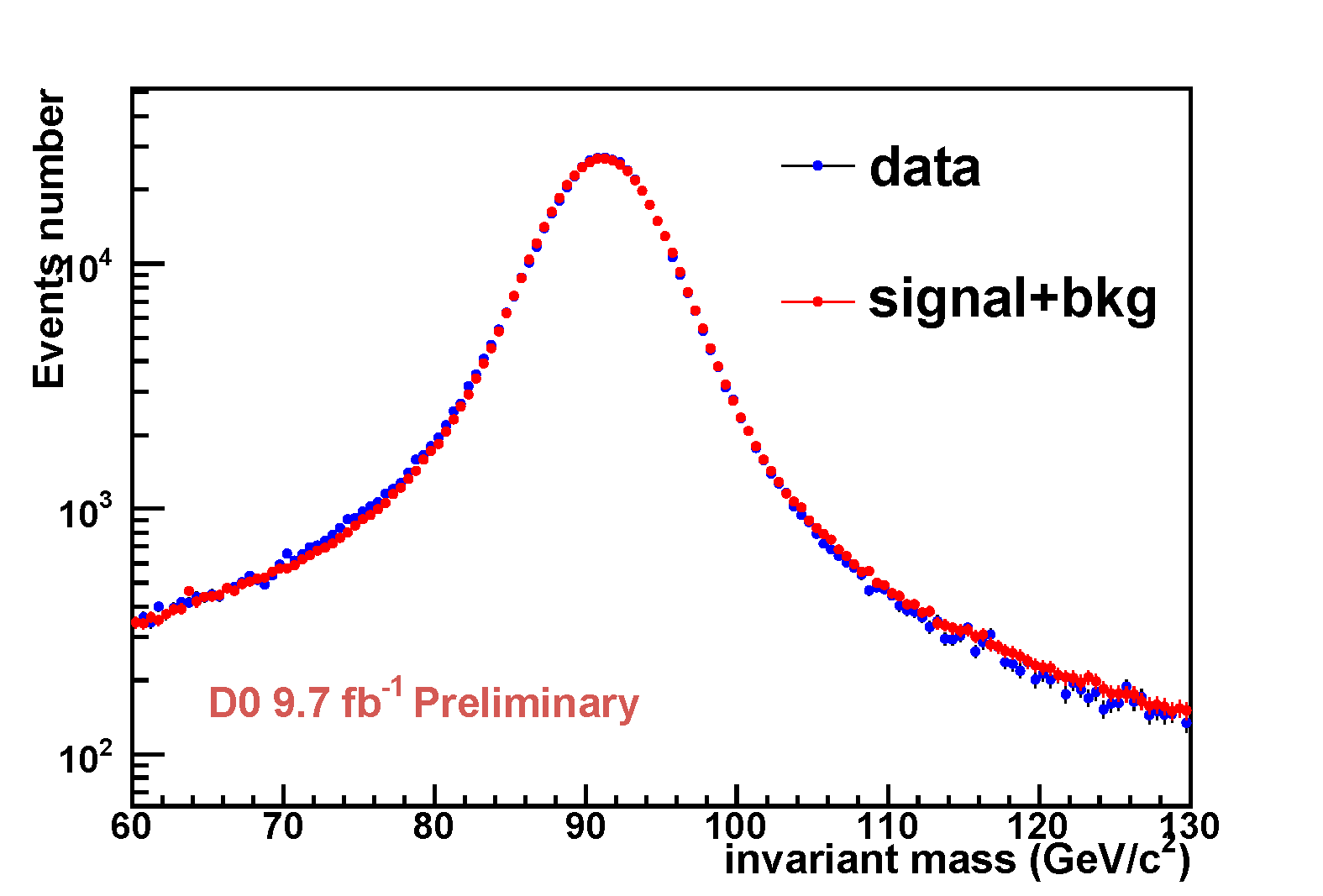}
\hfill
\includegraphics[width=0.5\linewidth]{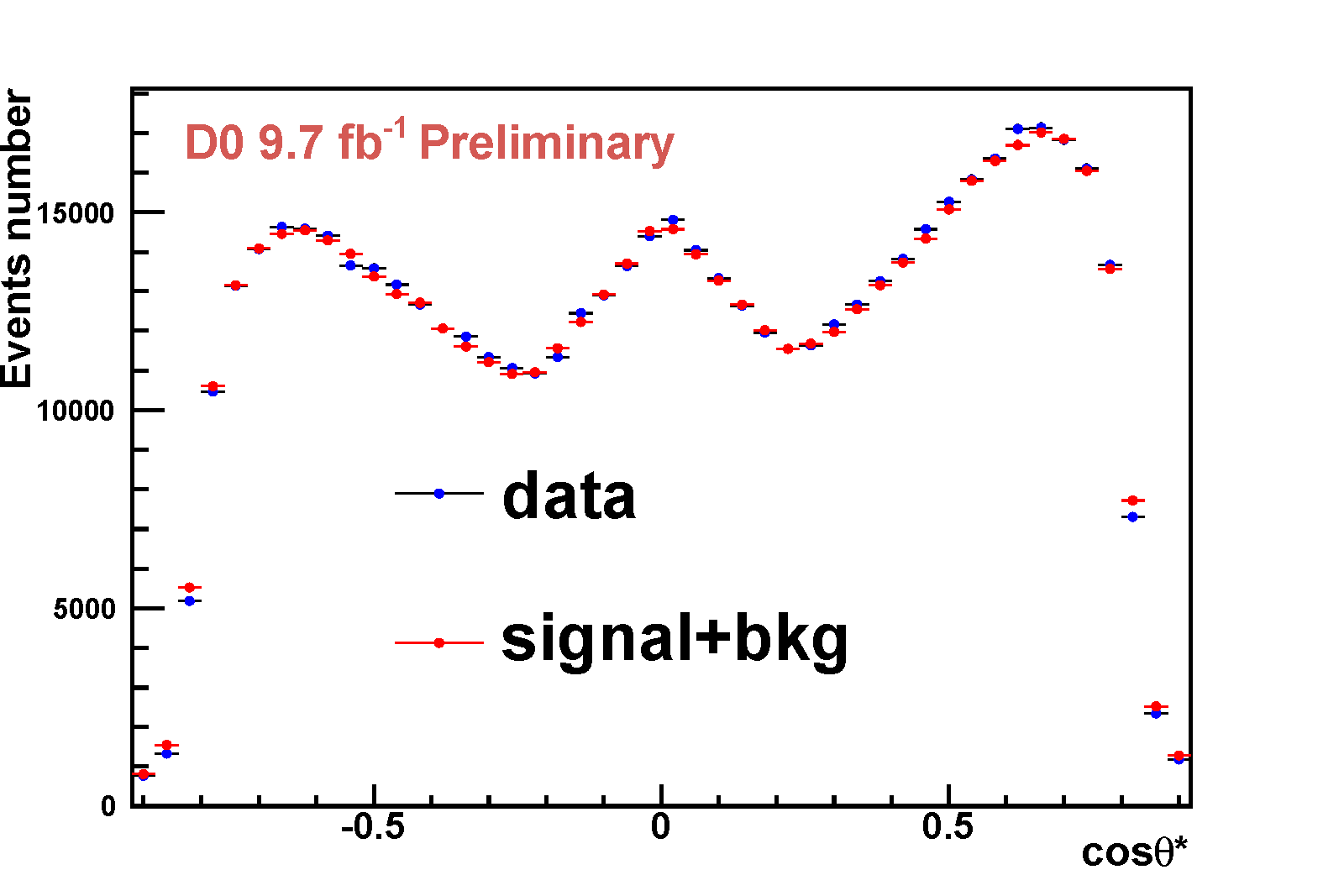}
\caption[]{D0 dielectron invariant mass (left) and Collins-Soper frame $\cost$ 
(right) distributions.}
\label{fig:D0dist}
\end{figure}

Electron reconstruction in the calorimeters depends critically on the elctron 
energy calibration.  The global energy scale modeling used in the previous D0 
analysis is inadequate for the current measurement because shape dependencies of
the different detector responses over the extended acceptance regions are too 
great.  A new method was developed that corrects energy as a function of 
instantaneous luminosity ($L_{inst}$) first, and then as a function of detector 
pseudorapidity ($\eta_{det}$).  This procedure scales the $Z$ mass peak to the 
LEP value of 91.1875 GeV in each ($L_{inst},\eta_{det}$) bin.  Calibrations are
performed separately for data and MC.  After calibration, the mass peak 
dependence on $L_{inst}$ is negligible, and the dependence on $\eta_{det}$ is 
reduced from 2 GeV to 100 MeV for data and 10 MeV for MC.  The improvement of
$\eta_{det}$ dependence in data is illustrated in Fig.~\ref{fig:escale}.  
Energy resolution and efficiency corrections are applied, and reweightings 
particularly as functions of $L_{inst}$ and vertex z are performed.  Final
distributions of $M$ and $\cost$ for data and MC are displayed in 
Fig.~\ref{fig:D0dist}.

\begin{figure}
\centerline{\includegraphics[width=0.6\linewidth]{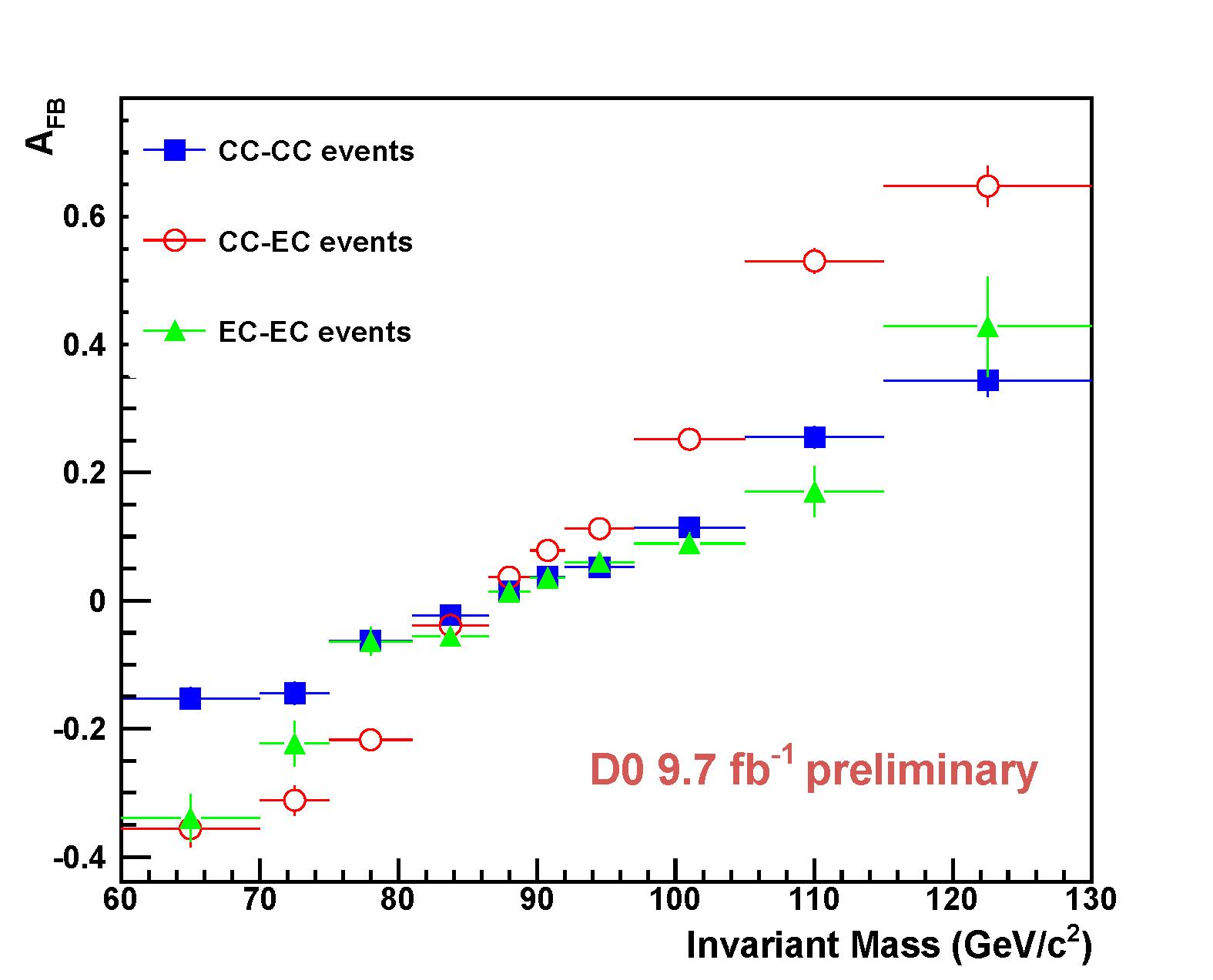}}
\caption[]{D0 $\afb$ distributions for all three dielectron event topolgies.}
\label{fig:D0_Afb}
\end{figure}

The $\sineff$ extraction is performed separately for subsamples based on location
of the two electrons (CC-CC, CC-EC, EC-EC), and on run period (RunIIa: 
1.1 fb$^{-1}$ low $L_{inst}$, RunIIb: 8.6 fb$^{-1}$ high $L_{inst}$).  The full 
dataset distributions of $\afb$ for all three event topologies can be seen in 
Fig.~\ref{fig:D0_Afb}.  Determination of $\sineff$ is made through $\chi^2$
comparisons of these distributions with MC templates generated at different 
values of $\sinw$ by rewieghting the generator level ($M_{Z/\gamma^*},\cost$)
distribution from the default MC ($\sinw = 0.232$).  The PDF uncertainty is the
dominant systematic uncertainty in this analysis, with smaller contributions from
energy scale and smearing, charge and electron misidentification, and 
backgrounds.  The extracted value of $\sinw$ is 
$0.23098\pm0.00042_{stat}\pm0.00014_{syst}\pm0.00029_{PDF}$.  In a SM context 
with on-shell renormalization scheme and \textsc{resbos} EBA correction, this corresponds 
to $\sineff = 0.23106\pm0.00053$.  This is the most precise determination of 
$\sineff$ from a hadron collider and from light quark couplings.

\section{Conclusions}

\begin{figure}
\centerline{\includegraphics[width=0.7\linewidth]{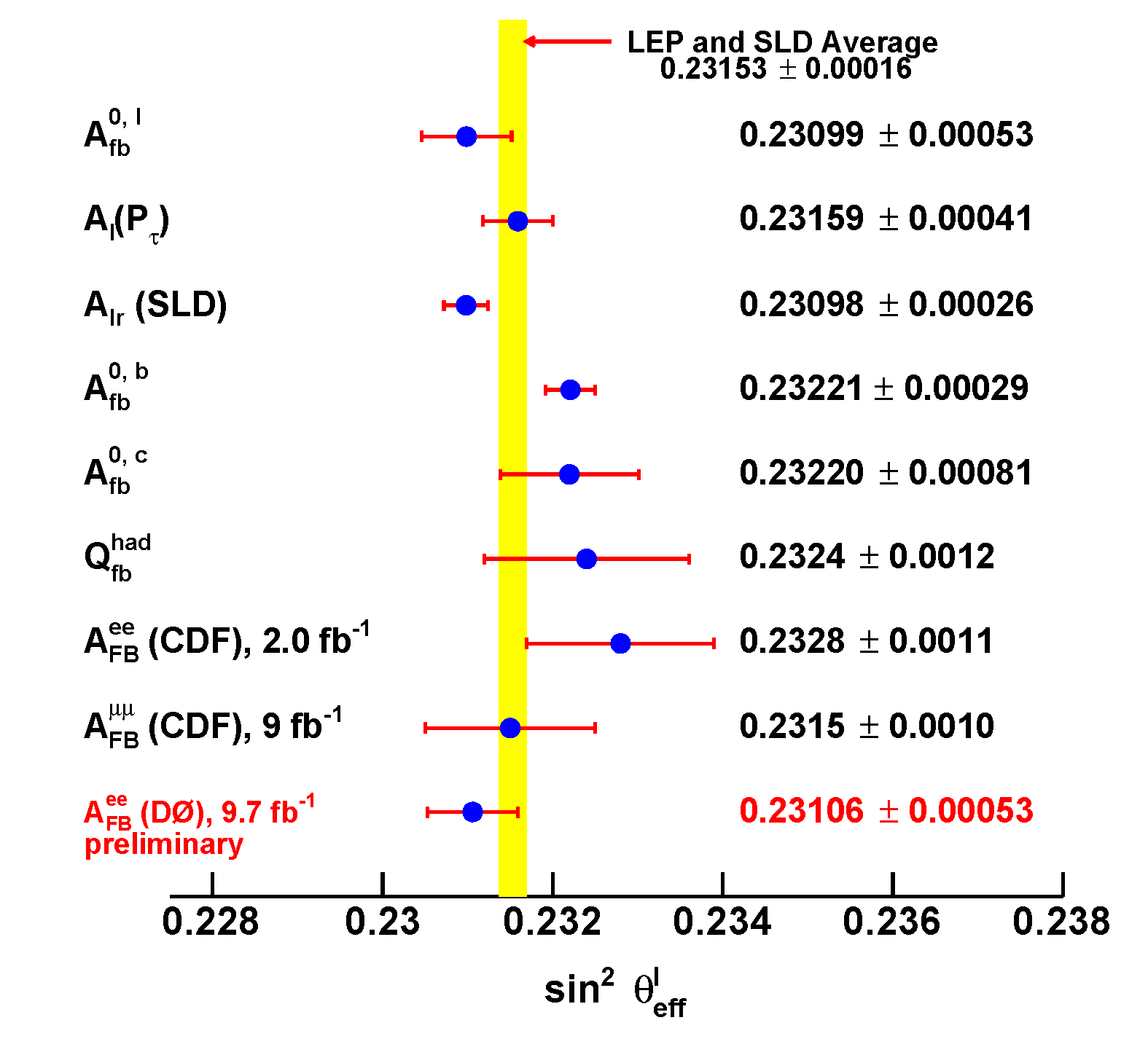}}
\caption[]{Comparison of Tevatron $\sineff$ results with those from LEP and SLD
experiments.}
\label{fig:sin2ThetaW}
\end{figure}

We report two new measurements of $\sineff$ from the forward-backward asymmetry
of Drell-Yan pairs at the Tevatron.  CDF finds a value of 
$\sineff = 0.2315\pm0.0010$ from studies of dimuon events, and D0 obtains a 
preliminary result of $\sineff = 0.23106\pm0.00053$ from a sample of dielectron 
events.  Both results are compared to and shown to be consistent with previous 
determinations of $\sineff$ from the LEP and SLD Collaborations in Fig.~\ref{fig:sin2ThetaW}.~\cite{lepsld} 

\end{document}